\documentstyle[sprocl,epsfig,amssymb]{article}

\def\be{\begin{equation}}
\def\ee{\end{equation}}
\def\bea{\begin{eqnarray}}
\def\eea{\end{eqnarray}}

\def\ifmath#1{\relax\ifmmode #1\else $#1$\fi}%
\def\rD{\ifmath{{\mathrm{D}}}}

\def\rF{\ifmath{{\mathrm{F}}}}

\def\rT{\ifmath{{\mathrm{T}}}}
\def\rX{\ifmath{{\mathrm{X}}}}

\def\rp{\ifmath{{\mathrm{p}}}}
\def\rq{\ifmath{{\mathrm{q}}}}
\def\rW{\ifmath{{\mathrm{W}}}}

\def\min{\ifmath{{\mathrm{min}}}}

\def\in{\ifmath{{\mathrm{in}}}}
\def\out{\ifmath{{\mathrm{out}}}}

\def\JJ{\ifmath{{\mathrm{JJ}}}}
\def\jet{\ifmath{{\mathrm{jet}}}}
\def\JET{\ifmath{{\mathrm{JET}}}}
\def\Bj{\ifmath{{\mathrm{Bj}}}}
\def\SD{\ifmath{{\mathrm{SD}}}}
\def\DPE{\ifmath{{\mathrm{DPE}}}}
\def\ND{\ifmath{{\mathrm{ND}}}}

\begin{document}
\title{DIFFRACTIVE STRUCTURE FUNCTIONS AT THE TEVATRON}

\author{MICHAEL G. ALBROW}

\address{Fermilab, Wilson Road,\\ Batavia, IL 60510\\E-mail: albrow@fnal.gov}

\author{For the CDF Collaboration\\~~}

\maketitle\abstracts{
We have studied events with a high-$x_\rF$ antiproton and two central jets
with $E_\rT>7$ GeV in CDF, in $\rp\bar{\rp}$ collisions at 
$\sqrt{s} = 1800$ GeV.
From the di-jet kinematics we derive the diffractive structure
function of the antiproton.
We also find an excess of events with a rapidity gap at least 3.5 units wide
in the proton direction, which we interpret as di-jet production in
double pomeron exchange events. We find non-factorization between our
single diffractive results and (a) diffraction in ep at HERA
(b) double pomeron exchange.}

\section{Introduction}
 There are predominantly three classes of hard diffractive events
at the Tevatron ($\rp\bar{\rp}$ collisions at $\sqrt{s} = 1.8$ TeV).
In {\em Single Diffraction}, SD, the p or $\bar{\rp}$ emerge with
$p_{\out} = (1-\xi)p_{\in}$ with $\xi<0.1$ (ideally $\xi \lessapprox 0.05$).
It has small 4-momentum transfer squared $t=(p^4_{\in}-p^4_{\out})^2$.
In hard {\em Double Diffraction}, DD, between a high $E_\rT$ jet at large
positive $\eta$ and one at large negative $\eta$ there is a 
rapidity {\em gap} of
at least 3 units, and in hard {\em Double Pomeron Exchange},
DI$\!\!\!$PE, high $E_\rT$
jets in the central region have rapidity gaps on both sides, with a leading
p and $\bar{\rp}$. In SD, we can consider an exchanged entity, which we
can call the pomeron I$\!\!\!$P for convenience (though there can be some other
reggeons I$\!\!\!$R exchanged too), with momentum fraction $\xi$. If a pair
of high $E_\rT$ jets are produced, with pseudorapidities $\eta$,
we can calculate the Bjorken-$x$, $x_{\Bj}$, 
of the scattering partons from:\footnote{Strictly speaking, the formula 
should have true rapidity $y$ not pseudorapidity $\eta$.}
\begin{equation}
x_{\Bj} = \frac{1}{\sqrt{s}}\sum_{JETS }E_\rT e^{\pm\eta}\ .
\end{equation}
If the interacting partons have fractional momenta $\beta$ of I$\!\!\!$P,
then $\beta = \frac{x_{\Bj}}{\xi}$. If all the hadrons $i$
in the event are measured,
with $E_\rT^i$ and $\eta_i$, then $\xi$ can also be determined from:
\begin{equation}
\xi = \frac{1}{\sqrt{s}}\sum_{particles}E_\rT^i e^{\pm\eta_i}\ ,
\end{equation}
where the +(--) sign is for diffractively scattered $\bar{\rp}(\rp)$.
(The p have large positive $\eta$.) We define a {\em diffractive structure
function} ${\cal F}$$(\xi,t,x_{\Bj},Q^2)$. In pomeron language, we measure
the parton distribution function in the I$\!\!\!$P. In structure function
language, we measure the ${\cal F}$ of the p (or $\bar{\rp}$)
when we have a large gap or a leading p (or $\bar{\rp}$).

In Run I we measured, using the rapidity gap technique, the
diffractive production of di-jets, $W$, high $E_\rT$ b-jets, and J$/\psi$,
and Jet-Gap-Jet JGJ events. Then, for the last two months of the run, we
installed scintillating fiber tracking hodoscopes in roman pots to detect
diffractively scattered $\bar{\rp}$. We measured $\bar{\rp}$JJ (SD) \cite{apjj}
and $\bar{\rp}$JJG (DI$\!\!\!$PE) \cite{pjjg} events.
The central detectors consist of tracking chambers in a 1.5T solenoidal
field, surrounded by sampling electromagnetic and hadronic calorimetry
used here for measuring particles and jets.

Our studies of diffraction using rapidity gaps can be summarized by saying
that for all hard processes (at $\sqrt{s}$ = 1.8 TeV) {\em approximately}
1\% are diffractive. Comparing diffractive W, b-jet and di-jet production
we can estimate that at the relevant $Q^2 \approx M_\rW^2$ the
gluon fraction in the I$\!\!\!$P is 0.54$\pm$0.15. W are made by $\rq\bar{\rq}$
annihilation, b-jets by gg interactions and generic di-jets by a mixture.
The soft I$\!\!\!$P at $Q^2 \approx 0$ may still be dominated by gluons.
Another result is that if one uses a POMPYT
Monte Carlo tuned to ep diffractive data and assumes factorization,
one predicts a factor $\approx$ 5 too much hard diffraction at the Tevatron.
This is an indication of the breakdown of factorization, implying that one
cannot treat hard diffractive processes as the {\em emission} of a pomeron
with a universally-defined parton distribution function (Ingelman-Schlein
model \cite{INGE}).

Although in this talk I concentrate on CDF results, D\O\  have
presented \cite{abbo} a study of events with forward rapidity gaps and either
forward or central di-jets at both $\sqrt{s} = 630$ and 1800 GeV. They
compared ratios of these conditions with Monte Carlo predictions
with hard, soft or flat g-distributions in the pomeron. None of these choices
fits the data. It is possible to compare the data to a prediction with a flat
distribution and extract an effective distribution; also to use
equ. (1) and (2) to estimate $x_p$ and $\xi$. However, what one
really wants to do is to tag the I$\!\!\!$P by measuring the
quasi-elastic (anti-)proton, so that one can measure the diffractive structure
functions directly.

\section{Single Diffraction with measured $\bar{\rp}$}
Near the end of Run I we installed small (2 cm $\times$ 2 cm)
scintillating fiber x,y hodoscopes in three roman pots 56 m downstream
from CDF. Antiprotons are measured after passing
through 81.6 Tm of dipole field. In one low luminosity run
at $\sqrt{s}$ = 1.8 TeV, 3.1M events were
collected which after clean up (single high quality central vertex,
single clean pot track with $0.035<\xi<0.095, |t|<1.0$ GeV$^2$)
reduces to 1,639k events. Of these, 30.4k = 1.86\% have at least 2 jets
with $E_\rT>7$ GeV. For comparison we took 342k non-diffractive events,
of which 32.6k = 10.9\% have at least two 7 GeV jets. The main reason
why ND events are 6 times richer in jets than SD events is the higher available
energy (1800 GeV cf $\sqrt{\xi s}$ = 330 - 550 GeV). The fraction of
diffractive triggers that have di-jets rises with $\xi$ from 0.013 to 0.022.
Interestingly there is no dependence on $|t|$ from $|t|_{\min}$ to 3 GeV$^2$,
\begin{equation}
t_{\min} = 2 \left[m_p^2 - (E_{\in}E_{\out} - p_{\in}p_{\out})\right]\ .
\end{equation}
The $E_\rT^{\JET}$ spectrum extends to 35 GeV, and is steeper than the ND
jet spectrum. The jets are boosted towards the proton direction 
(positive rapidity) by about 1 unit, on average.

We can use the jet kinematics to measure Bjorken-$x$, $x_{\Bj}$, of
the colliding partons, see equ. (1).
A 3rd jet is included if it has $E_\rT> 5$ GeV. Let the ratio of SD:ND jets
be $R(\frac{SD}{ND})(x_{\bar{\rp}},\xi)$ and the ND di-jet effective structure
function be:
\begin{equation}
F_{jj}(x_{\bar{\rp}}) = x_{\bar{\rp}}\left[g(x_{\bar{\rp}}) + \frac{4}{9}
\sum (q_i(x_{\bar{\rp}})+\bar{\rq}_i(x_{\bar{\rp}}))\right]\ .
\end{equation}
For $F_{jj}(x_{\bar{\rp}})$ we use the GRV98LO structure function. Then
\begin{equation}
F^\rD_{jj}(x_{\bar{\rp}},\xi) = R(\frac{\SD}{\ND})(x_{\bar{\rp}},\xi) \times
F_{jj}(x_{\bar{\rp}})
\end{equation}
and through the change of variables $\beta= \frac{x_{\bar{\rp}}}{\xi}$ 
we can derive $F^\rD_{jj}(\beta,\xi)$.

The ratio $R(\frac{\SD}{\ND})(x_{\bar{\rp}})$ for different $\xi$ is shown
in Fig.1 vs $x_{\Bj}$(antiproton) = $x_{\bar{\rp}}$.
It is not flat, showing that when a lower
$x_{\Bj}$ parton gives rise to the jets it is more likely (than a higher
$x_{\Bj}$ parton) to be a
diffractive event. The middle part of the distribution, not affected by
kinematic boundaries, fits a power law:
\begin{equation}
R(\frac{\SD}{\ND})(x_{\bar{\rp}}) = (6.1 \pm 0.1)\times 10^{-3}
\times (x_{\bar{\rp}}/0.0065)^{-0.45\pm0.02}\ .
\end{equation}

\vspace{4mm}
\centerline{
\epsfig{file=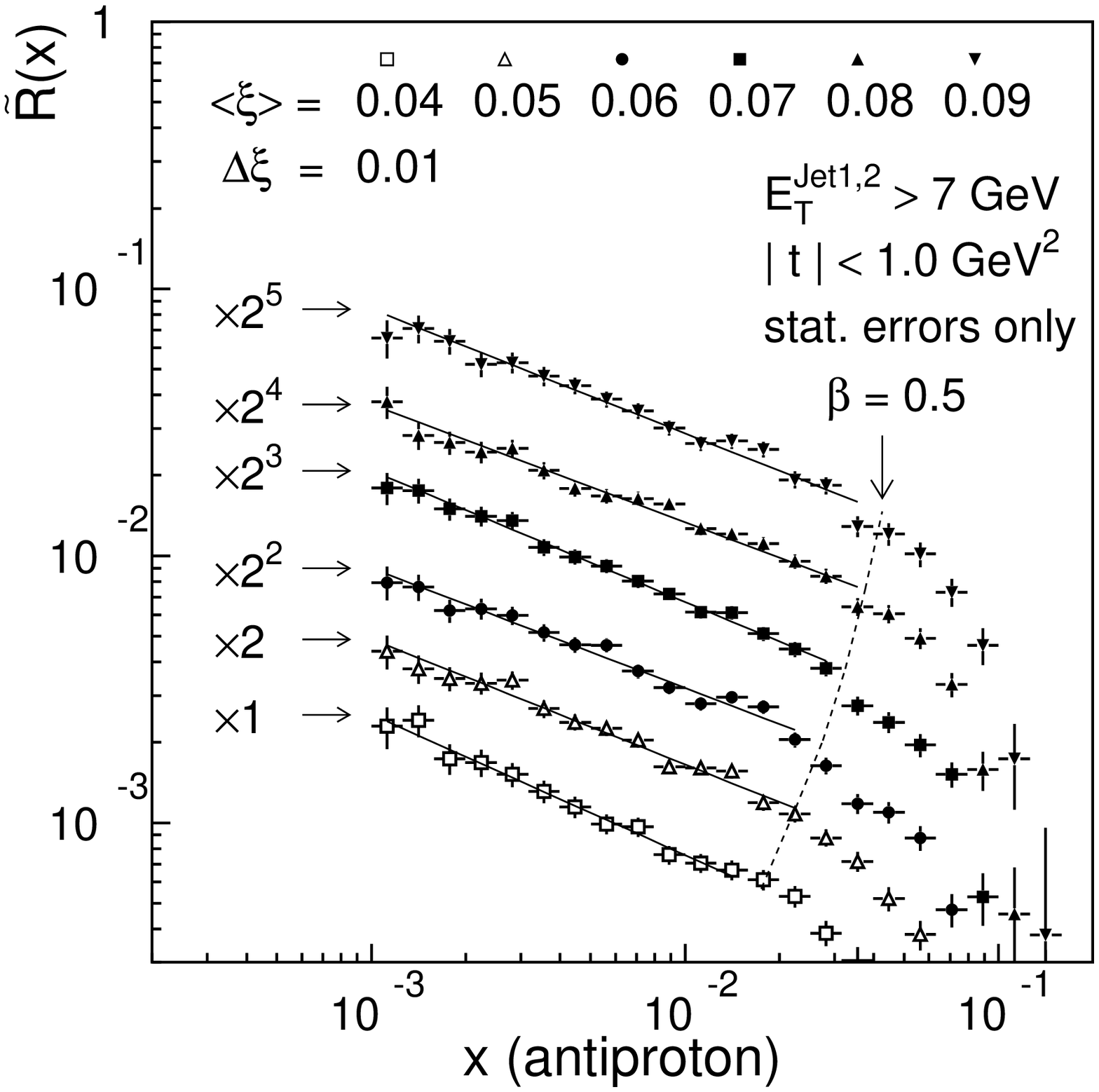,width=7.2cm}}
\vspace{1mm}
{\small\baselineskip=10pt\noindent
Figure 1. Ratio of diffractive to non-diffractive di-jet event rates as 
a function of $x_{\Bj}$. Different $\xi$ regions are plotted separately
and displaced for clarity.\par}
\vspace{4mm}

\centerline{
\epsfig{file=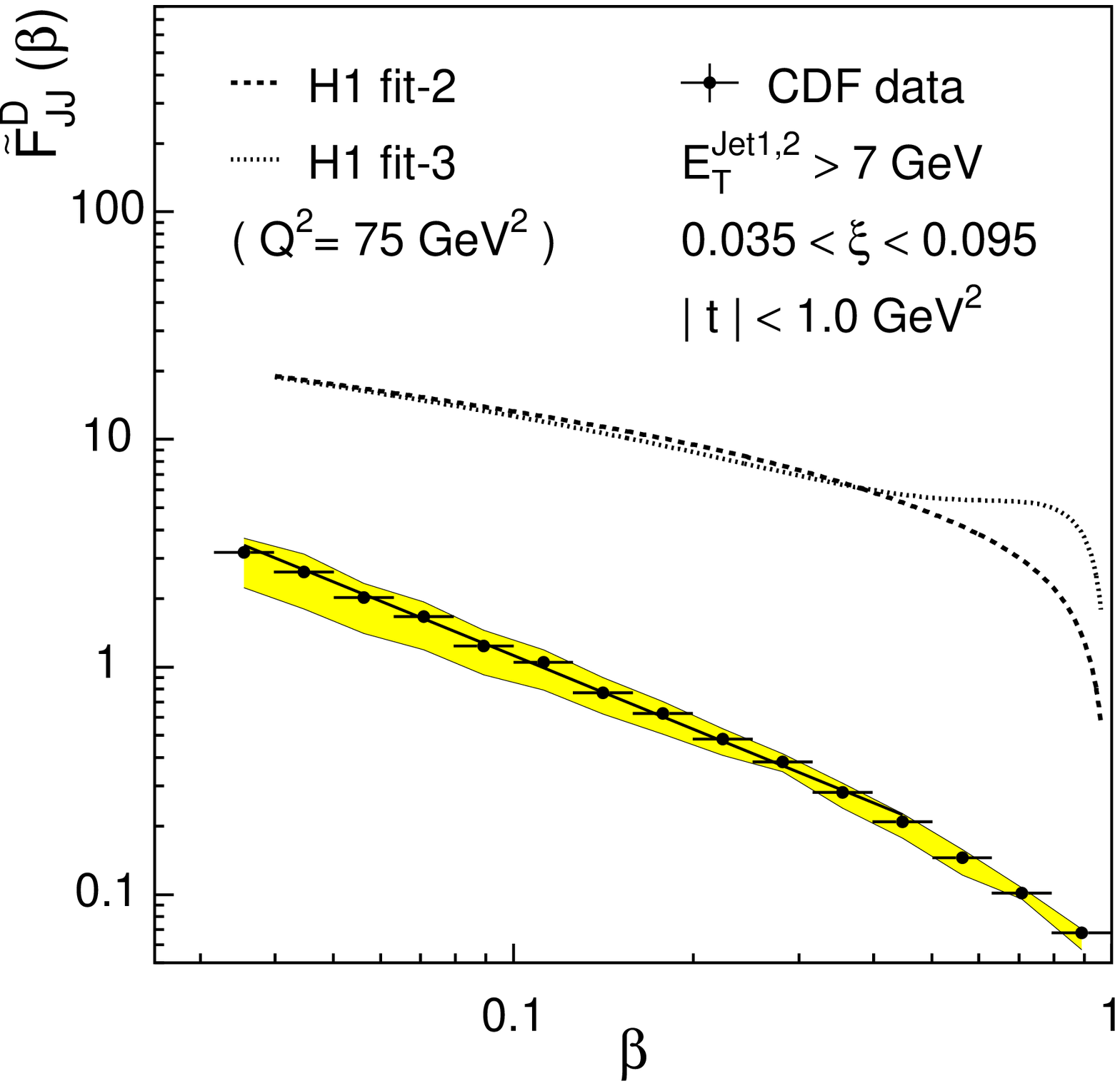,width=7.2cm}}
\vspace{1mm}
{\small\baselineskip=10pt\noindent
Figure 2. Data $\beta$ distribution (points) compared with
expectations from diffractive deep inelastic scattering (H1).
The dashed(dotted) lines are from H1 fit 2(3).\par}
\vspace{4mm}

When we change variables and plot $F_{jj}^\rD(\beta)$ vs $\beta$ we find (see
Fig.2) a falling distribution with no sign of a $\beta=1$ peak (``superhard
pomeron") and normalization a factor $>10$ lower than the H1 fits (ep).
This is showing a breakdown of (ep $\leftrightarrow$ pp) factorization.
The $x_{\Bj}$(antiproton) and $\beta$ distributions appear to be independent
of $\xi$ from 0.035 to 0.095; we have a factorizing form
for $\frac{10^{-3}}{\xi}<\beta<0.5$:
\begin{equation}
F^\rD_{jj}(\beta,\xi) = C.\beta^{-1.04\pm0.01}\xi^{-m}\ .
\end{equation}
At larger $\beta$, $m$ rises from $\approx$ 0.85
at $\beta$ = 0.03 to $\approx$ 1.2 at $\beta$ = 1,
similar to the behavior observed at H1 and ZEUS.

\section{Double Pomeron Exchange}
We now turn to double pomeron exchange, which provides another
test of factorization. We take the SD di-jet events and look for
rapidity gaps ($\Delta\eta >3.5$) on the outgoing p-side, by counting 
hits in the region $2.4<\eta<5.9$. We find 132 events with 0 hits, 
while an extrapolation
of the bulk of the multiplicity distribution gives 14 events, so the signal
is very significant with a S:N = 10:1. (This is with both jet $E_\rT>7$ GeV;
with a 10 GeV cut we still have 17 events.)
The mean $<\xi_{\bar{\rp}}>$ is slightly larger 
(within the range 0.035 - 0.095)
for these DI$\!\!\!$P events than for the SD events. As well as measuring
$\xi_{\bar{\rp}}$ with the pot track, we can estimate it from the sum over
all hadrons, see equation (2).
By comparing the two we find a correction factor of 1.7 (not all particles
are detected and/or well measured). Then we can estimate the $\xi_\rp$
of the unseen proton as
\begin{equation}
\xi_\rp = 1.7 \times \frac{1}{\sqrt{s}}\sum E_\rT^i e^{+\eta_i}\ .
\end{equation}
Most of the events have $0.01<\xi_\rp<0.03$ and for definitiveness
we make that cut. The $E_\rT$ spectrum of the jets is like that in SD,
but the statistics are poor above 12 GeV. The $\eta$ of the di-jet
tends to be negative (towards the $\bar{\rp}$) as one expects from
$\xi_\rp - \xi_{\bar{\rp}}$.
In fact:
\begin{equation}
\Delta\eta \approx \ln\xi_{\bar{\rp}} - \ln\xi_\rp\ .
\end{equation}
The $\Delta\phi$ distribution between the leading jets
is slightly more peaked at $180^\circ$ than in SD (and SD more than ND).
The fraction of all the central mass carried by the two leading jets,
$R^{\JJ}_\rX$, is a broad distribution with no sign of a peak near 1, 
which might be expected from a non-factorizing DPE diagram.

Just as before we could plot $R(\frac{\SD}{\ND})(x_{\bar{\rp}})$, now we can
plot $R(\frac{\DPE}{\SD})(x_\rp)$ in the same range of $x_{\Bj}$. If we had
factorization among these processes we would expect these quantities to
be equal; in fact the latter is a factor $\approx$5 higher. This does not seem
to be due to the different $\xi$ ranges of the p and $\bar{\rp}$, 
as we saw before
no $\xi$-dependence in $R(\frac{\SD}{\ND})(x_{\bar{\rp}})$.

We can quote a cross section for $0.01<\xi_\rp<0.03$,
$0.035<\xi_{\bar{\rp}}<0.095$, $-4.2<\eta^{\jet1,2}<+2.4$,
$E_\rT^{\jet1,2}>7(10)$ GeV of 43.6$\pm$22 (3.4$\pm$2.2) nb, where 
the error is essentially all systematic resulting from the difficulty 
of measuring such low $E_\rT$
jets. Removing the $\xi_\rp$ cut to gain statistics, at
95\% c.l. $<$ 8.5\% of the 7 GeV di-jets are exclusively produced,
i.e. with $R^{\JJ}_\rX \approx$ 1. This is much smaller than a prediction
of $\approx$ 1 $\mu$b.\cite{bere}

In conclusion, we measured non-factorizing
effects in hard diffraction. The ``pomeron" does not have a unique
structure function; it depends on the environment (ep, p$\bar{\rp}$(SD),
p$\bar{\rp}$(DPE)). In hard diffraction, a more realistic picture than 
the Ingelman-Schlein
model is probably one where a hard scatter takes place (on a very short
space-time scale) and before or after (on a much longer space-time scale)
soft gluons are exchanged which can create rapidity gaps and can leave
the $\rp,\bar{\rp}$ or both in their ground state. These soft
color interactions are quite different when the interaction involves
an $e_{\in},e_{\out},\gamma$ than when it does not.

\section{Acknowledgements}
I thank Dino Goulianos, Albert de Roeck and the organizers of ISMD
2000 for their invitation, and the Fermilab staff and the staffs of
CDF institutions for their contributions. This was a supplementary
program (E876) within CDF, and additional support from the
U.S. Department of Energy and the Ministry of Education, Science and
Culture of Japan is gratefully acknowledged.

\end{document}